\documentclass[preprint,12pt]{elsarticle}
\usepackage{graphicx}
\usepackage{amssymb}
\usepackage{amsmath}

\journal{Nuclear Physics A}

\begin{document}

\begin{frontmatter}

\title{$K^- N$ amplitudes below threshold constrained by multinucleon 
absorption} 

\author[a]{E.~Friedman\corref{cor1}}
\author[a]{A.~Gal}
\cortext[cor1]{Corresponding author: E. Friedman, elifried@cc.huji.ac.il}
\address[a]{Racah Institute of Physics, The Hebrew University, 91904 
Jerusalem, Israel}

\begin{abstract}
Six widely different subthreshold $K^- N$ scattering amplitudes obtained in 
SU(3) chiral-model EFT approaches by fitting to low-energy and threshold data 
are employed in optical-potential studies of kaonic atoms. Phenomenological 
terms representing $K^-$ multinucleon interactions are added to the 
EFT-inspired single-nucleon part of the $K^-$-nucleus optical potential in 
order to obtain good fits to kaonic-atom strong-interaction level shifts and 
widths across the periodic table. Introducing as a further constraint the 
fractions of single-nucleon $K^-$ absorption at rest from old bubble-chamber 
experiments, it is found that only two of the models considered here reproduce 
these absorption fractions. Within these two models, the interplay between 
single-nucleon and multinucleon $K^-$ interactions explains features 
observed previously with fully phenomenological optical potentials. Radial 
sensitivities of kaonic atom observables are re-examined, and remarks are made 
on the role of `subthreshold kinematics' in absorption-at-rest calculations. 

\end{abstract}


\end{frontmatter}

\section{Introduction}
\label{sec:intro}

Several SU(3) chiral-model EFT approaches to the $\bar K$-nucleon interaction 
have been presented in recent years, demonstrating good agreement with each 
other at threshold and above, where they are constrained by some $K^-p$ 
low-energy scattering and reaction data. At threshold, the precise measurement 
of strong-interaction level shift and width of the 1$s$ state in the kaonic 
atom of hydrogen by the SIDDHARTA collaboration~\cite{Baz11,Baz12} provides 
an important constraint on the $K^-$p interaction, in addition to three 
previously known threshold branching ratios. However, as shown recently 
\cite{CMM16,Kam16} the resulting $K^- N$ scattering amplitudes appear to 
be strongly model dependent when extrapolating to energies below threshold 
and also with respect to their pole content. Moreover, for the $K^- n$ 
amplitude different models predict widely different results both below 
threshold and above. 

An algorithm for constructing self-consistently the density dependent 
single-nucleon part $V^{(1)}_{K^-}(\rho)$ of the $K^-$-nucleus optical 
potential $V_{K^-}(\rho)$ from energy dependent $\bar K N$ amplitudes was 
presented by Ciepl\'{y} et al. \cite{CFG11a,CFG11b}. For the predominantly 
attractive $\bar K N$ interaction, the $\bar K N$ energies involved in kaonic 
atoms are below threshold, leading to a well defined in-medium `subthreshold 
kinematics' scheme, as practised in our recent analyses of kaonic atom data 
\cite{FGa12,FGa13}. For lack of accepted microscopic theory, this approach 
cannot be applied to the multinucleon part $V^{(2)}_{K^-}(\rho)$ of 
$V_{K^-}(\rho)$, which usually consists of phenomenological terms devised to 
achieve agreement with experiment, as reviewed in Refs.~\cite{BFG97,FGa07}. 

In the present work we consider six different $K^- N$ amplitudes derived 
in several EFT SU(3) chiral-model approaches to $\cal S$=$-$1 meson-baryon 
interactions and studied recently for their pole content \cite{CMM16}. 
Augmented by phenomenological amplitudes that could be identified with 
multinucleon processes, these six single-nucleon model amplitudes are used 
to describe strong interaction observables in kaonic atoms. These observables 
include for the first time some old experimental results from bubble chambers 
on absorption at rest of $K^-$ mesons on nuclei \cite{DOK68,MGT71,VSW77}. 
Specifically, we calculate the {\it fraction} of absorptions on a single 
nucleon out of all absorptions at rest in kaonic atoms, and compare with 
these experimental results in order to place a new constraint on the 
$K^-N$ model-amplitudes input. While this topic was discussed for 
$^{12}$C \cite{YHi07} and in nuclear matter \cite{SYJ12}, the
present study is the first realistic kaonic atoms calculation that considers 
single-nucleon absorption fractions across the periodic table. 

The paper is organized as follows. Section~\ref{sec:method} introduces the 
$K^-N$ model-amplitudes input and the in-medium `subthreshold kinematics' 
methodology as applied to kaonic atoms, followed by the topic of 
single-nucleon fractions of absorption at rest. Section~\ref{sec:results} 
provides results, showing that only two of the six sets of $K^- N$ amplitudes 
are acceptable when fractions of single-nucleon vs. multinucleon absorption 
are considered in kaonic atoms studies. The two acceptable sets of amplitudes 
turn out to be quite similar, and in section~\ref{sec:discuss} we discuss 
in detail the interplay between the single-nucleon amplitudes and the full 
optical potential based on these two sets of amplitudes. Radial sensitivities 
of kaonic atom observables are re-examined, and remarks are made on the role 
that `subthreshold kinematics' plays in absorption-at-rest calculations. 
A brief summary is presented in section~\ref{sec:summ}.

\begin{figure}[!ht]
\begin{center}
\includegraphics[height=70mm,width=0.80\textwidth]{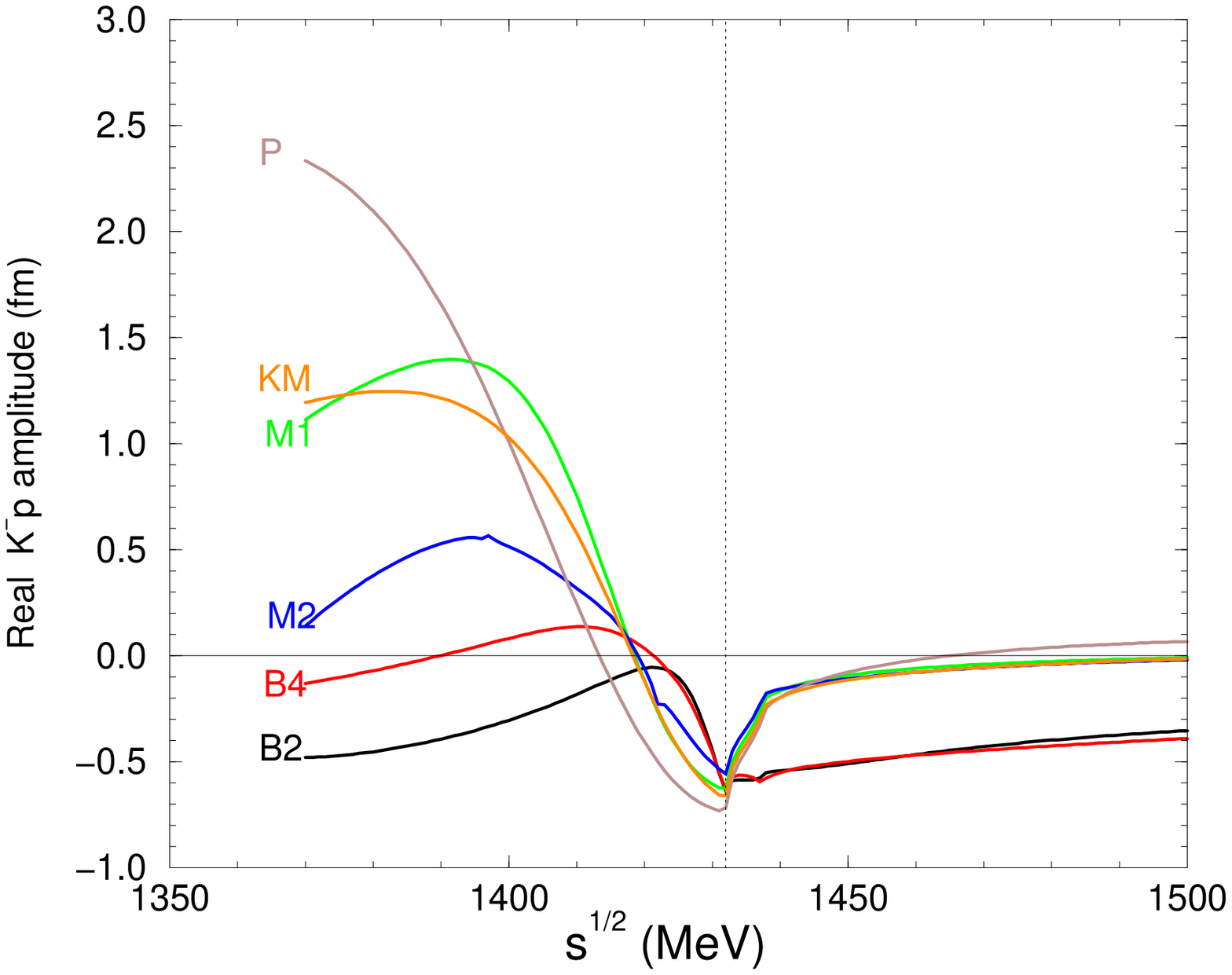}
\includegraphics[height=70mm,width=0.80\textwidth]{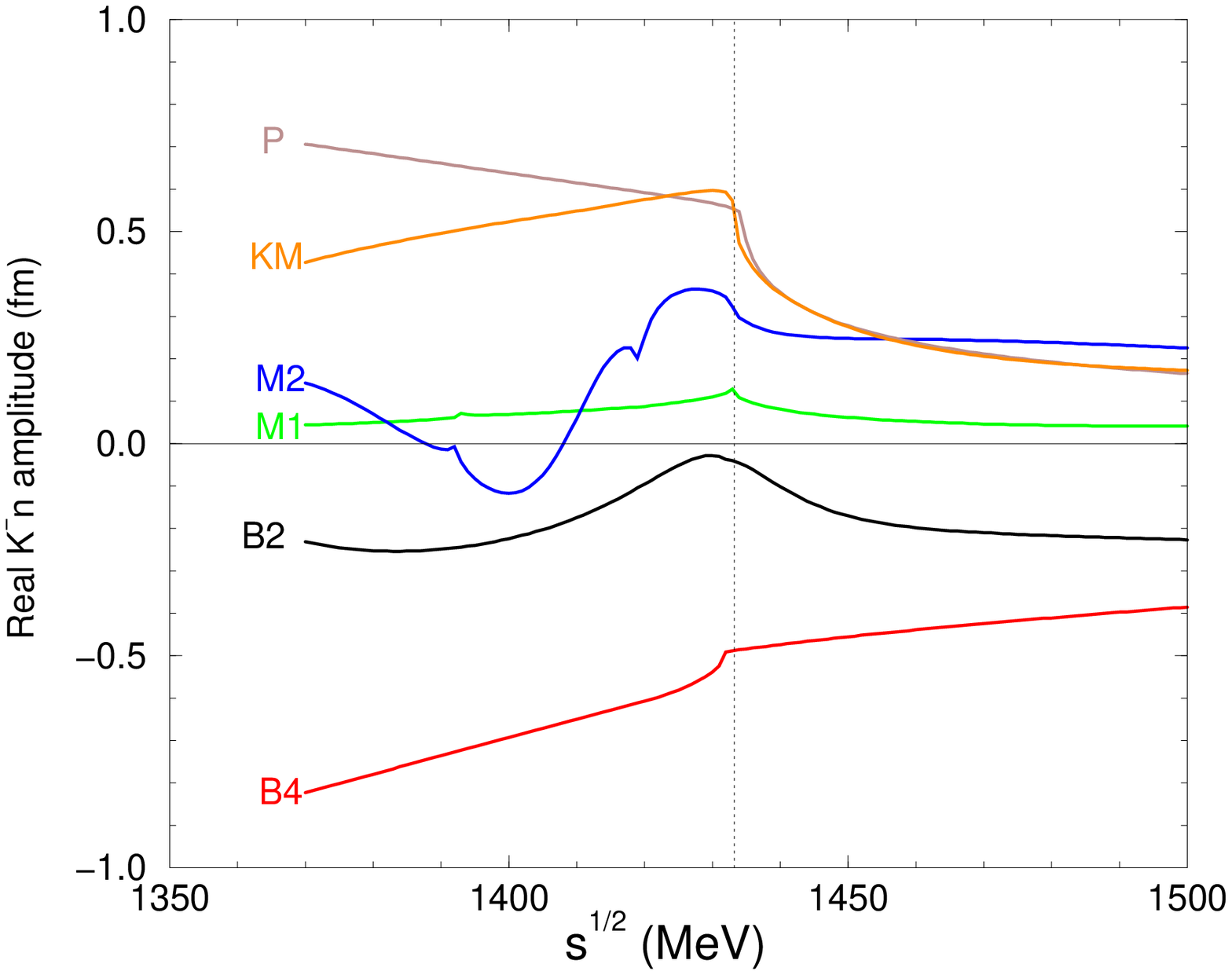}
\caption{Color online. Real part of six $K^- N$ $s$-wave amplitudes used 
in the present work, taken from chiral SU(3) meson-baryon coupled-channel 
dynamical models discussed in \cite{CMM16} and denoted there by KM 
for Kyoto-Munich \cite{IHW12}, P for Prague \cite{CSm12}, M1 and M2 
for Murcia \cite{GOl13}, and B1 and B2 for Bonn \cite{MMe15}. Vertical 
dotted lines indicate threshold energies.}
\label{fig:Kpreal}
\end{center}
\end{figure}

\section{Input and methodology} 
\label{sec:method} 

\subsection{Input $K^- N$ amplitudes} 
\label{subsec:input} 

\begin{figure}[!ht]
\begin{center}
\includegraphics[height=70mm,width=0.80\textwidth]{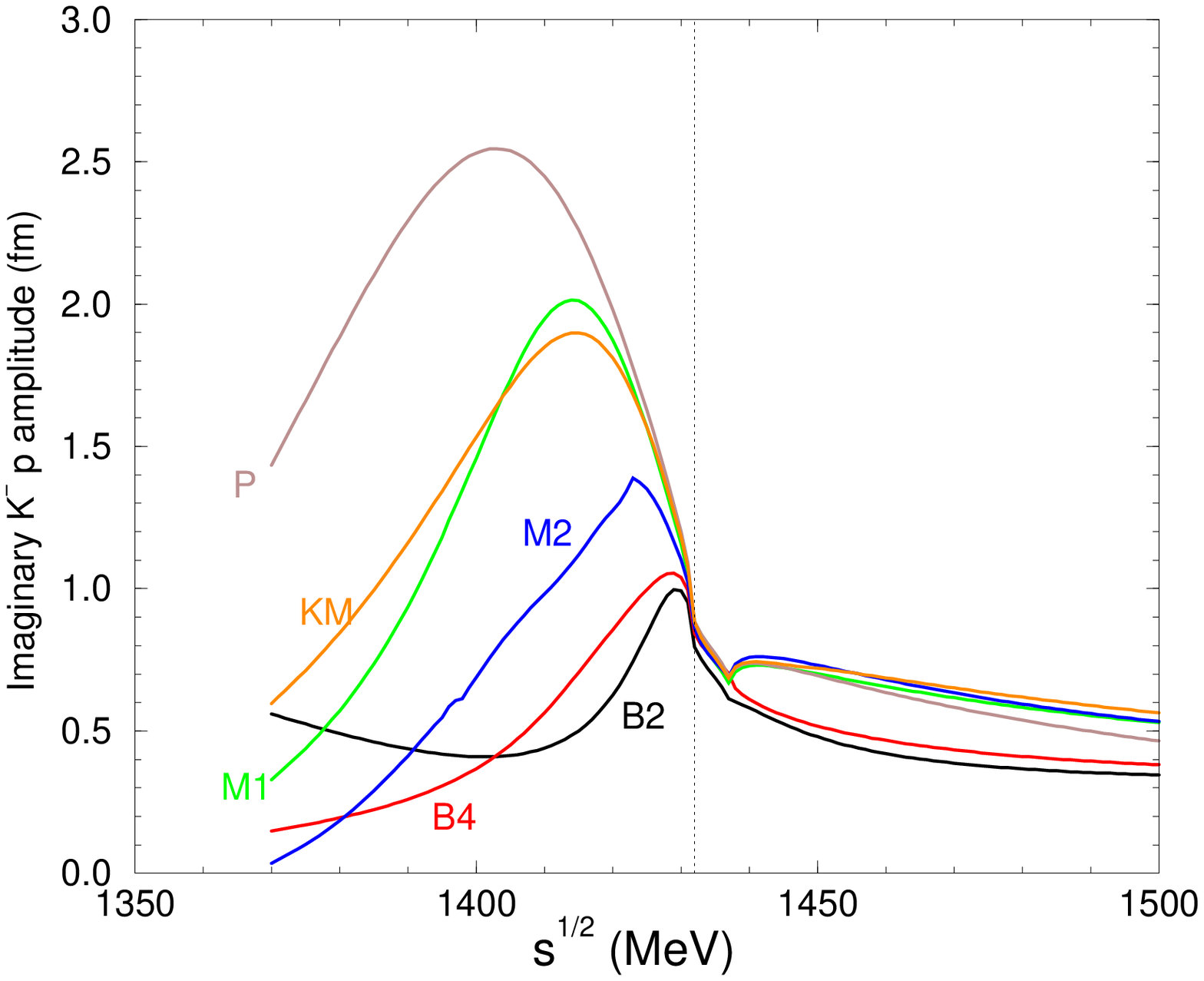}
\includegraphics[height=70mm,width=0.80\textwidth]{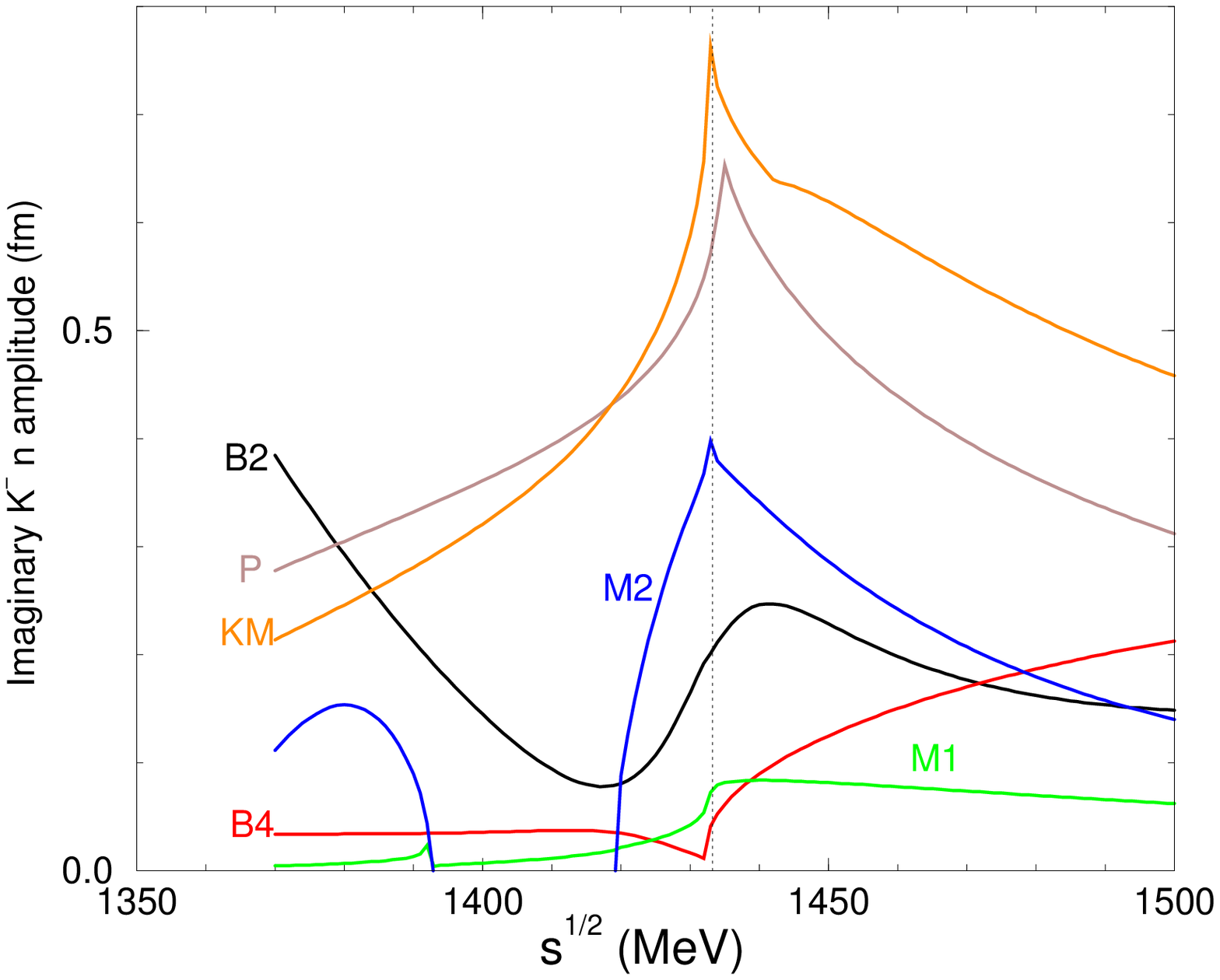}
\caption{Color online. Imaginary part of six $K^- N$ $s$-wave amplitudes 
used in the present work. Notations follow those in Fig.~\ref{fig:Kpreal}. 
Vertical dotted lines indicate threshold energies.}
\label{fig:Kpimag}
\end{center}
\end{figure}

The $K^- N$ amplitudes considered in the present work have been derived in 
various strangeness $\cal S$=$-$1 chiral SU(3) meson-baryon coupled-channel 
dynamical model approaches at next-to-leading order (NLO). Our specific 
choice follows that by Ciepl\'{y} et al. \cite{CMM16} in their recent 
pole-content study of six such amplitudes generated in four different 
models, \cite{IHW12,CSm12,GOl13,MMe15} in chronological order. At NLO, 
the meson-baryon interaction consists of contact terms, parametrized by 
low-energy constants (LEC) which are fitted to low-energy $K^-p$ scattering 
and reaction cross sections, to three precisely determined threshold 
branching ratios, and also to the kaonic hydrogen energy shift and width 
as measured recently at the DA$\Phi$NE $e^+e^-$ collider in Frascati by 
the SIDDHARTA Collaboration \cite{Baz11,Baz12}. These NLO fitted LEC are 
added to the leading-order (LO) Weinberg-Tomozawa (WT) term plus the Born 
graphs contributions. For a recent review, see Ref.~\cite{Kam16}. The six 
$K^- N$ $s$-wave center-of-mass (c.m.) amplitudes considered in the present 
work are plotted in Figs.~\ref{fig:Kpreal} (real parts) and \ref{fig:Kpimag} 
(imaginary parts) as a function of the c.m. energy $\sqrt{s}$ below and above 
threshold for both $K^-p$ (upper panels) and $K^-n$ (lower panels). 

For $K^-p$, all the models agree at threshold, and three out of the four 
models also agree above threshold where they were essentially fitted to 
data. The Bonn model amplitudes, however, differ from the other ones above 
threshold, apparently because higher partial waves were allowed in the fitting 
procedure. All the amplitudes differ appreciably when going to subthreshold 
energies. For $K^-n$, there are substantial differences between any two of the 
plotted amplitudes over the full energy range shown. For M2, in particular, 
the imaginary amplitude is not even positive definite. Thus, below threshold 
where one derives the input $K^-N$ amplitudes for kaonic-atom calculations 
(see next subsection) all six amplitudes shown in Figs.~\ref{fig:Kpreal} and 
\ref{fig:Kpimag} are different, providing thereby an interesting model 
dependence to explore. 

The Kyoto-Munich (KM) and Prague (P) $K^- N$ amplitudes were used already 
in our previous kaonic atom studies \cite{FGa12,FGa13}, yielding similar 
results to each other, but within a different context than the present one. 
These two amplitudes also happen to satisfy the absorption-at-rest constraint 
imposed in the present work. Interestingly, the KM and P NLO amplitudes 
are only moderately different from their LO WT variants, in contrast to the 
Murcia (M) and Bonn (B) amplitudes with sizable NLO contributions. As noted 
in Ref.~\cite{CMM16}, the available low-energy $K^-p$ data do not exclude 
large NLO contributions, although good reproduction of these data was reached 
already at the LO level. In this respect, the Prague and Kyoto-Munich models 
are the more conservative among the models surveyed here.

\subsection{Subthreshold kinematics}
\label{subsec:kinema}

The model underlying the subthreshold energy algorithm adopts the Mandelstam 
variable $s = (E_{K^-} + E_N)^2 -(\vec p _{K^-} + \vec p_N)^2$ as the 
argument transforming free-space to in-medium $K^- N$ amplitudes, 
where both the $K^-$ and the nucleon variables are determined independently 
by the respective environment of a $K^-$ atom and a nucleus. Consequently, 
unlike in the two-body c.m. system, here $\vec p _{K^-} + \vec p_N$ does not 
vanish, and one gets to a good approximation 
$(\vec p _{K^-}+\vec p_N)^2\rightarrow p _{K^-}^2+p_N^2$ 
upon averaging over angles. Energies are given by 
\begin{equation} 
\label{eq:energies} 
E_{K^-} =m_{K^-} -B_{K^-},~~~~ E_N=m_N-B_N, 
\end{equation} 
where $m$ denotes masses and $B$ denotes binding energies. For the $K^-$ 
momentum we substitute locally 
\begin{equation}
\label{eq:locpi}
\frac{p_{K^-}^2}{2 m_{K^-}} = -B_{K^-} - {\rm Re}~V_{K^-} -V_c \, , 
\end{equation} 
where $V_{K^-}$ is the $K^-$-nucleus optical potential and $V_c$ is the $K^-$ 
Coulomb potential due to the finite-size nuclear charge distribution. 
For the nucleon momentum $p_N$ we adopt the Fermi gas model (FGM), yielding 
in the local density approximation 
\begin{equation} 
\label{eq:Fermi} 
\frac{p_N^2}{2 m_N} = T_N\; (\rho / {\bar\rho})^{2/3}, 
\end{equation} 
where $\rho$ is the local density, $\bar\rho$ is the average nuclear density 
and $T_N$ is the average nucleon kinetic energy which assumes the value 
23~MeV in the FGM. 

Defining $\delta \sqrt s =\sqrt s -E_{\rm th}$ with $E_{\rm th}=m_{K^-}+m_N$, 
and applying the minimal substitution requirement \cite{ET82,KKW03} 
$E \rightarrow E-V_c$, then to first order in $B/E_{\rm th}$ and 
$(p/E_{\rm th})^2$ one gets 
\begin{equation} 
\delta \sqrt s = 
-B_N\rho/{\bar \rho} 
-\beta_N [T_N(\rho/\bar{\rho})^{2/3}+B_{K^-}\rho/\rho_0+ 
V_c (\rho/\rho_0)^{1/3}] 
+\beta_{K^-}{\rm Re}~V_{K^-}, 
\label{eq:SCmodifMS} 
\end{equation} 
with $\beta_N=m_N/(m_N+m_{K^-}),~\beta_{K^-} = m_{K^-}/(m_N+m_{K^-})$, 
and $\rho_0=0.17$~fm$^{-3}$. Following previous applications 
\cite{CFG11a,CFG11b,FGa12,FGa13} an average binding energy value of 
$B_N=8.5$~MeV is used. The specific $\rho/\rho_0$ and $\rho/{\bar \rho}$ 
forms of density dependence ensure that $\delta\sqrt{s}\rightarrow 0$ when 
$\rho \rightarrow 0$~\cite{FGa13} in accordance with the low-density limit.

\subsection{In-medium amplitudes}
\label{subsec:medium}

Eq.~(\ref{eq:SCmodifMS}) defines energy as function of a local density 
where the optical potential is part of the argument. In the `$t \rho $' 
approximation (see below) the potential depends on the $K^- N$ amplitudes 
that in turn depend on the potential through the energy $\sqrt{s}$. 
Therefore an iterative solution of Eq.~(\ref{eq:SCmodifMS}) is required. 
Experience shows that convergence is achieved after 4-6 iterations, 
yielding the energy $\sqrt{s}$ as function of density $\rho$ to serve 
as the argument for calculating the amplitudes. 

Final steps towards a calculation of a `$t \rho $' optical potential are 
the transformation from the $K^- N$ system to the $K^-$-nucleus system and 
incorporating Pauli correlations, which are the longest-range correlations 
in the nuclear medium, from Ref.~\cite{WRW97}. The `$t \rho $' in-medium 
single-nucleon optical potential $V^{(1)}_{K^-}(\rho)$, including Pauli 
correlations, is given by \cite{FGa13} 
\begin{equation} 
2{\mu_K}V^{(1)}_{K^-}(\rho)=-4\pi \left[ \frac{(2{\tilde f}_{K^-p}-
{\tilde f}_{K^-n})\:\frac{1}{2}\rho_p}{1+\frac{1}{4}\xi_k(\rho){\tilde f}_0
\rho(r)}+\frac{{\tilde f}_{K^-n}(\frac{1}{2}\rho_p+\rho_n)}
{1+\frac{1}{4}\xi_k(\rho){\tilde f}_1\rho(r)}\right] , 
\label{eq:wrw14} 
\end{equation} 
where $\mu _K$ is the $K^-$-nucleus reduced mass and the in-medium 
$K^- N$ scattering amplitudes in the $K^-$-nuclear c.m. frame, 
${\tilde{f}}_{K^-N}(\rho)$, are  related kinematically to the in-medium 
$K^- N$ c.m. amplitudes $f_{K^-N}(\rho)$ plotted in Figs.~\ref{fig:Kpreal} 
and~\ref{fig:Kpimag} by ${\tilde{f}}_{K^-N}(\rho)=(1+\frac{A-1}{A}\frac{\mu_K}
{m_N}){f}_{K^-N}(\rho)$. These amplitudes, 
in view of Eq.~(\ref{eq:SCmodifMS}), are now density dependent. 
The Pauli correlation factor $\xi_k(\rho)$ is defined by \cite{WRW97}  
\begin{equation} 
\xi_k(\rho)=\frac{9\pi}{k_F^2}\left( 4\int_0^{\infty}{\frac{dr}{r}\,
\exp({\rm i}kr)\,j_1^2(k_F r)} \right) \; ,
\label{eq:wrw25} 
\end{equation} 
with $k=[(E_{K^-}-i \Gamma /2)^2-m_K^2]^{1/2}$ and where $\Gamma$ is the 
width of the particular kaonic atom state. The Fermi momentum is given by 
$k_F=(3{\pi}^2\rho/2)^{1/3}$; $\tilde f_0$ and $\tilde f_1$ are the isospin 
0 and 1 combinations, respectively, of the $K^-N$ amplitudes; and $\rho_p$ 
and $\rho_n$ are proton and neutron densities, respectively. 
Denoting the integral in Eq.~(\ref{eq:wrw25}) by $I_k(\rho)$, 
it can be evaluated analytically, 
\begin{equation} 
4 I_k (\rho)=1-\frac{q^2}{6}+\frac{q^2}{4}(2+\frac{q^2}{6})\;{\rm ln}
(1+\frac{4}{q^2})-\frac{4}{3}q\;(\frac{\pi}{2}-{\rm arctg}(q/2)) , 
\label{eq:integ} 
\end{equation} 
with $q=-ik/k_F$. This expression corrects Eq. (A5) of Ref.~\cite{FGa13}. 
The density dependence of $\xi_k(\rho)\rho/4$ in the denominators of 
Eq.~(\ref{eq:wrw14}) reduces at threshold to $\xi_{k=0}(\rho)\rho/4=
(3k_F/2\pi)\propto \rho^{1/3}$, thereby giving rise to a leading 
$\rho^{4/3}$ Pauli-correlation contribution to the single-nucleon `$t\rho$' 
potential \cite{WRW97}  

The single-nucleon optical potential $V^{(1)}_{K^-}$ is often augmented by 
a phenomenological isoscalar potential $V^{(2)}_{K^-}$ of the form 
\begin{equation} 
2\mu_K V^{(2)}_{K^-}(\rho)=-4\pi \, B\,(\frac{\rho}{\rho_0})^{\alpha} 
\,\rho \, , 
\label{eq:V2} 
\end{equation} 
with a complex strength $B$ and a positive exponent $\alpha$, representing 
$K^-$ {\it multinucleon} processes. For $\alpha=1$, the $\rho ^2$ dependence 
of $V^{(2)}_{K^-}(\rho)$ agrees with that traditionally associated in mesic 
atoms with multinucleon meson interactions, motivated primarily by absorption 
on two nucleons \cite{EEr66,SMa83}. Although other effects as well could 
induce a $\rho ^2$ density dependence in leading order, e.g. self-energy 
(SE) insertions considered recently in Refs.~\cite{CFG11a,CFG11b,FGa12}, 
in this work we do not treat separately such $\rho ^2$ contributions. 
The added $V^{(2)}_{K^-}(\rho)$ vanishes faster than $\rho$ as $\rho 
\rightarrow$~0, and the in-medium $K^- N$ amplitude given by the square 
bracket of Eq.~(\ref{eq:wrw14}) for $V^{(1)}_{K^-}(\rho)$ agrees in this 
limit with that for free nucleons at threshold, consistently with the low 
density limit. Although $V^{(2)}_{K^-}(\rho)$ is introduced independently 
of $V^{(1)}_{K^-}(\rho)$, it affects implicitly the density dependence 
of the in-medium $V^{(1)}_{K^-}(\rho)$, the latter being determined 
by a self consistent application of $\delta \sqrt{s}(\rho)$, 
Eq.~(\ref{eq:SCmodifMS}), in which the {\it full} optical potential 
$V_{K^-}(\rho)=V^{(1)}_{K^-}(\rho)+V^{(2)}_{K^-}(\rho)$ appears.

\subsection{Absorption at rest}
\label{subsec:absorption}

Branching ratios of $K^-$ absorption at rest on protons and on several nuclear 
species have been measured many decades ago with nuclear emulsions and with 
bubble chambers (BC). However, since the absorption fractions reported from 
the latter method are quoted with considerably smaller uncertainties than 
those of the former, we use here exclusively the results of three such BC 
experiments \cite{DOK68,MGT71,VSW77}. Davis et al. \cite{DOK68} quote 
0.26$\pm$0.03 for the multinucleon capture fraction for $K^-$ on a mixture 
of C, F and Br. Moulder et al. \cite{MGT71} quote 0.28$\pm$0.03 for that 
fraction on Ne, and the fraction for C is given by Vander Velde-Wilquet et al. 
\cite{VSW77} as 0.19$\pm$0.03. We therefore adopt as a best estimate of an 
experimental $K^-$ multinucleon absorption-at-rest fraction an average value 
of 0.25$\pm$0.05 for C and heavier nuclei. This may be linked with strong 
interaction level widths in kaonic atoms. 

The level width $\Gamma $ which is usually obtained from the complex 
eigenvalue $E_{K^-}-i\Gamma/2$ when solving the Klein-Gordon equation with 
an optical potential $V_{K^-}$ \cite{FGa07}, is also related to the imaginary 
part of the potential by the overlap integral of ${\rm Im}\,V_{K^-}$ and 
$|\psi|^2$, 
\begin{equation} 
\label{eq:width} 
\Gamma~=-2~\frac{\int {\rm Im}\,V_{K^-}~|\psi|^2~d{\vec r}} 
{\int [1-(B_{K^-}+V_{\rm C})/\mu_K]~|\psi|^2~d{\vec r}} \, , 
\end{equation} 
where $B_{K^-}$, $V_{\rm C}$ and $\mu_K$ are the $K^-$ binding energy, 
Coulomb potential and reduced mass, respectively, and $\psi$ is the $K^-$ wave 
function of the particular state concerned \cite{FOk13}. For a Schroedinger 
equation the denominator is just the $\psi$ normalization integral. 
When the optical potential $V_{K^-}$ is the sum of a single-nucleon part 
$V^{(1)}_{K^-}$ and a multinucleon part $V^{(2)}_{K^-}$, it is possible 
to apply Eq.~(\ref{eq:width}) to each part, and thereby calculate 
single-nucleon and multinucleon absorption fractions, separately 
for any nucleus and for any specific kaonic atom state. 

\begin{figure}[!ht] 
\begin{center} 
\includegraphics[height=90mm,width=0.80\textwidth]{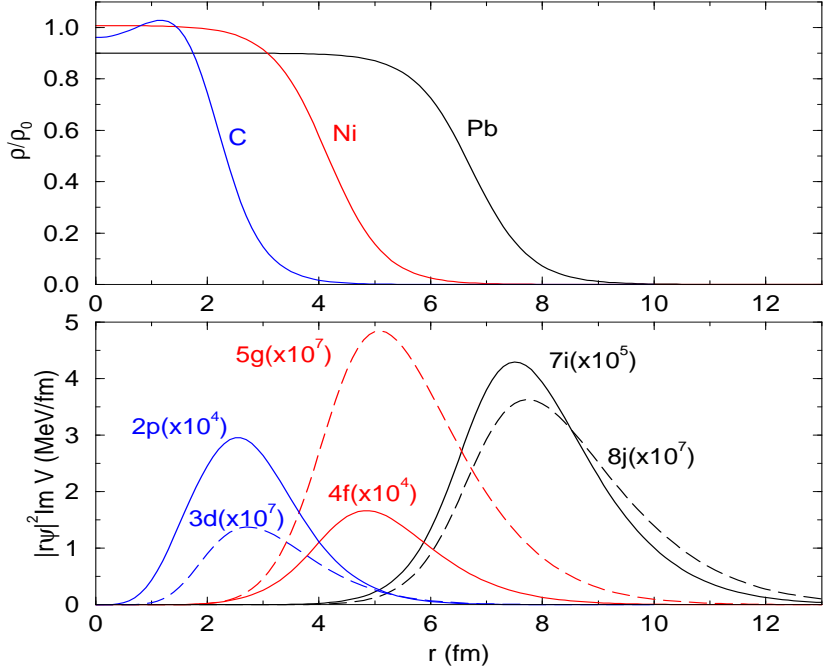} 
\caption{Color online. Lower panel: $|r\psi|^2{\rm Im}\,V_{K^-}$ 
for several kaonic atoms for a `lower' (solid curves)  and an `upper' 
(dashed curve) state. Top panel: the corresponding relative density 
distributions of the respective nuclei, with $\rho _0$=0.17~fm$^{-3}$.} 
\label{fig:overlaps} 
\end{center} 
\end{figure} 

Figure~\ref{fig:overlaps} shows on the lower panel the overlap between 
kaonic atom wave functions and the imaginary potential for C, 
Ni and Pb, separately for the `lower' and the `upper' levels \cite{FOk13}. 
The wave functions were calculated from a best-fit phenomenological 
potential denoted DD in Ref.~\cite{FGa07}. The upper panel of the figure 
shows the relative density distributions for these examples. The similarity 
between C, Ni and Pb is evident; the peak of the overlap for Ni and Pb 
is at densities near 15-20\% of nuclear matter density for the lower state 
and at 10-15\% for the upper state. These values are a little higher for C. 
In all cases the overlap extends inward to around 60\% of central density, 
as found in \cite{BFr07}. As the width of a level is proportional to the 
integral of the overlap for a kaonic atom state with the imaginary potential 
Eq.~(\ref{eq:width}), it is  expected following Fig.~\ref{fig:overlaps} 
that absorption at rest of $K^-$ mesons by nuclei will have very similar 
characteristic features for all targets from C and above along the periodic 
table. This is consistent with the limited data from BC experiments 
\cite{DOK68,MGT71,VSW77} that do not show any marked dependence of a 
multinucleon absorption fraction of 0.25$\pm$0.05 on the nuclear target. 
This value will serve here as a constraint on conventional analyses of 
strong-interaction observables in kaonic atoms. 

\section{Results}
\label{sec:results} 

\subsection{Fits to kaonic atom data}
\label{subsec:fits}

As in earlier publications \cite{BFG97,FGa07} the present work deals with 
global analyses of kaonic atoms, handling together experimental results 
along the periodic table in order to gain insight on average properties 
of the $K^-$-nucleus interaction near threshold. 

\begin{table}[htb] 
\caption{$\chi ^2$ values for 65 kaonic atoms data points using 
exclusively single-nucleon optical potentials $V^{(1)}_{K^-}(\rho)$, 
Eq.~(\ref{eq:wrw14}), based on six chiral-model $K^- N$ amplitudes. 
The density dependence of $V^{(1)}_{K^-}(\rho)$ was determined by solving 
Eq.~(\ref{eq:SCmodifMS}) self consistently.} 
\label{tab:input} 
\begin{center} 
\begin{tabular}{ccccccc} 
\hline 
model & B2 & B4 & M1 & M2 & P & KM \\ \hline
$\chi ^2$(65) &~ 1174~ & ~2358~& ~2544~& ~3548~& ~2300~ & ~1806~\\ \hline
\end{tabular}
\end{center}
\end{table}

Table \ref{tab:input} shows values of $\chi ^2$ for 65 kaonic atoms data 
points for an optical potential $V^{(1)}_{K^-}(\rho)$ constructed from 
Eq.~(\ref{eq:wrw14}) using the six single-nucleon amplitudes from 
Ciepl\'{y} et al.~\cite{CMM16} shown in Figs.~\ref{fig:Kpreal} and 
\ref{fig:Kpimag}. As is clear from this table, none of these amplitudes leads 
to agreement with experiment. This result is not new, as previous analyses 
have shown the need to include phenomenological terms motivated by $K^-$ 
multinucleon processes, $V^{(2)}_{K^-}$, in order to obtain agreement with 
experiment \cite{CFG11a,CFG11b}. Equally good fits of $\chi ^2$ per point of 
less than 2 could also be achieved with purely phenomenological forms for 
$V_{K^-}$ \cite{BFG97,FGa07}.

\begin{table}[htb]
\caption{Best-fit parameters and $\chi ^2$ values for 65 kaonic atoms data 
points, from $V^{(1)}_{K^-}$ optical potentials, Eq.~(\ref{eq:wrw14}), based
on six chiral-model $K^- N$ amplitudes, and supplemented by a $V^{(2)}_{K^-}$ 
potential, Eq.~(\ref{eq:V2}), representing multinucleon $K^-$ interactions. 
The density dependence of $V^{(1)}_{K^-}(\rho)$ was determined by solving 
Eq.~(\ref{eq:SCmodifMS}) with the full $V_{K^-}$ self consistently. 
Correlations between $\alpha$ and $B$ cause some of the errors to be large.} 
\label{tab:best} 
\begin{center} 
\begin{tabular}{ccccccc} 
\hline 
model & B2 & B4 & M1 & M2 & P & KM \\ \hline
$\alpha$ &0.29(10)&0.24(8)&0.24(16)&0.85(13)&1.4(1.2)&1.8(8) \\
Re$B$ (fm)&2.3$\pm$0.8&2.8$\pm$0.3&0.2$\pm$0.2&1.9$\pm$0.4&$-$1.2$\pm$0.3&
0.1$\pm$1.3 \\
Im$B$ (fm)&0.8$\pm$0.2&1.0$\pm$0.1&0.7$\pm$0.1&1.3$\pm$0.2&2.4$\pm$3.1&
3.3$\pm$1.1 \\
$\chi ^2$(65) & 111 & 105& 121& 108& 125 & 122\\ \hline
\end{tabular}
\end{center}
\end{table}

At the next stage, we add to the single-nucleon part $V^{(1)}_{K^-}$ 
a phenomenological part $V^{(2)}_{K^-}$, Eq.~(\ref{eq:V2}), in terms 
of an amplitude $B\,(\rho / \rho_0)^{\alpha}$ with three parameters: 
a complex strength $B$ and exponent $\alpha > 0$. Table~\ref{tab:best} 
shows results of fits to the data obtained by varying these three parameters. 
The density dependence of the single-nucleon part $V^{(1)}_{K^-}(\rho)$ of 
the full potential $V_{K^-}$ was determined by solving self consistently 
Eq.~(\ref{eq:SCmodifMS}) with the {\it full} potential $V_{K^-}$. 
It is seen that for each of the single-nucleon amplitudes that serve as 
the $V^{(1)}_{K^-}$ basis for the overall potential $V_{K^-}$, a good fit 
is obtained. In some cases the uncertainties are large due mostly to 
correlations between $B$ and $\alpha$. For the first three listed model 
amplitudes the values of $\alpha$ are significantly smaller than 1, whereas 
for the other three $\alpha$ is consistent with 1, meaning a $\rho ^2$ 
dependence of the potential. 

A comment on uncertainties is in order. Although dealing with 65 kaonic atom 
data points, only 3--4 parameters can be obtained by fitting to these data, 
with very different sensitivities for the exponent $\alpha$ and the other 
parameters. Once the single-nucleon part $V^{(1)}_{K^-}$ serves as a fixed 
part of the optical potential $V_{K^-}$, the exponent is hard to fit, 
with consequences for the other parameters. In the present analysis, 
the uncertainties were derived from the covariant matrix, indicating 
correlations between parameters. Most of the results in Table~\ref{tab:best} 
demonstrate this problem which is overcome by scanning over $\alpha$. 

\begin{table}[htb]
\caption{Same as in Table~\ref{tab:best}, except that $\alpha$ is kept fixed 
 at values compatible with its best-fit values.} 
\label{tab:fixed1} 
\begin{center} 
\begin{tabular}{ccccccc}
\hline 
model & B2 & B4 & M1 & M2 & P & KM \\ \hline
$\alpha$ &0.3&0.3&0.3&1.0&1.0&1.0 \\
Re$B$ (fm)&2.4$\pm$0.2&3.1$\pm$0.1&0.3$\pm$0.1&2.1$\pm$0.2&
$-$1.3$\pm$0.2&$-$0.9$\pm$0.2 \\
Im$B$ (fm)&0.8$\pm$0.1&0.8$\pm$0.1&0.8$\pm$0.1&1.2$\pm$0.2&
1.5$\pm$0.2&1.4$\pm$0.2 \\
$\chi ^2$(65) & 111 & 105& 121& 109& 125 & 123\\ \hline
\end{tabular}
\end{center}
\end{table}

Table~\ref{tab:fixed1} shows results of fits when values of $\alpha$ 
are kept fixed close to the best-fit values of Table~\ref{tab:best}. 
The $\chi ^2$ values are indeed as low as for the best fit when $\alpha $ 
too was varied. The last two columns (P and KM) show essentially 
the same additional term ($B$ and $\alpha$) within uncertainties. Inspecting 
Fig.~\ref{fig:Kpreal} and Fig.~\ref{fig:Kpimag} it is seen that there are 
indeed great similarities between the two models, except at energies 
lower than 30 to 40 MeV below threshold. Although it could be argued that 
the values of $\alpha $ for the first three listed amplitudes (B2, B4 and M1) 
are inconsistent with a plausible $\rho ^2$ dependence of the additional term, 
all six fits are essentially acceptable as fits to kaonic atom data. 
One needs additional constraints to narrow down the range of acceptable 
single-nucleon amplitudes.

Uncertainties in $\alpha$ were taken into account by calculating best-fit 
values of $B$ for  $\alpha$ between 1.0 and 2.5, thus 
making sure that these $(B, \alpha)$ sets fit well the kaonic atom data. 
This provides the starting point for discussing the single-nucleon 
absorption fractions.

\subsection{Fraction of single-nucleon absorptions}
\label{subsec:1Nfrac} 

Fractions of absorption on a single nucleon were calculated for all 
24 species of kaonic atoms included in the data base. These fractions 
were calculated from Eq.~(\ref{eq:width}) separately for the lower and for 
the upper levels, using the six in-medium chiral-model $K^- N$ amplitudes 
discussed above, supplemented by a phenomenological term 
$B(\rho /\rho_0)^\alpha$ as given in Table \ref{tab:fixed1}. 

\begin{figure}[!ht]
\begin{center}
\includegraphics[height=90mm,width=0.80\textwidth]{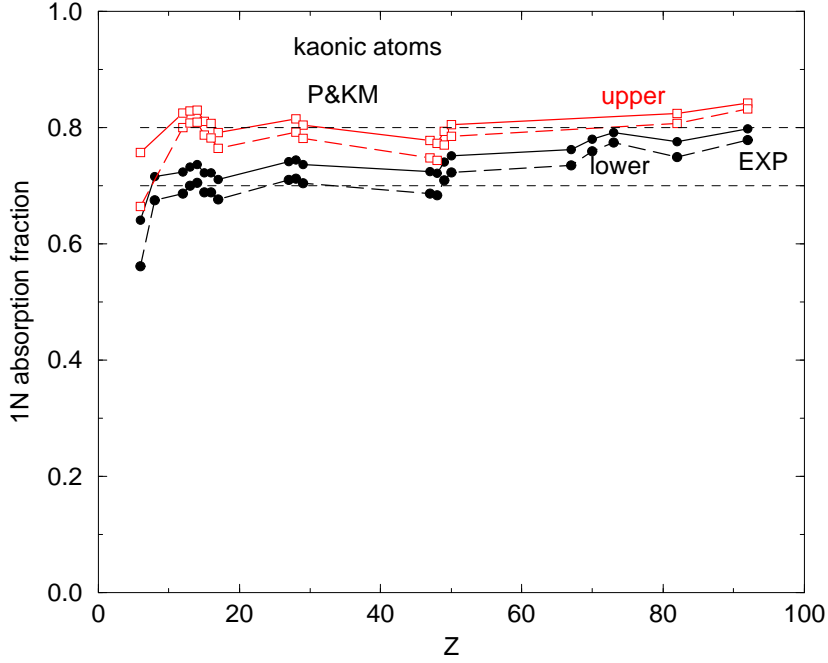}
\caption{Color online. Fractions of single-nucleon absorption for amplitudes 
P and KM: solid circles are for lower states, open squares for upper states, 
solid curves use $B$ values from Table~\ref{tab:fixed1} for $\alpha$=1, 
long-dashed curves use $B$ values for $\alpha$=2. Horizontal dashed 
lines mark the range of experimental values of the single-nucleon 
absorption fraction. Models P and KM are indistinguishable from each 
other on this figure. See text for discussion of uncertainties.} 
\label{fig:PKMfrac}
\end{center}
\end{figure}

Figure~\ref{fig:PKMfrac} shows calculated fractions of single-nucleon 
absorption when using the chiral-model input amplitudes P and KM. The 
two horizontal dashed lines represent our best estimate for the range 
of single-nucleon absorption fractions, 0.75$\pm$0.05, as deduced from 
BC studies \cite{DOK68,MGT71,VSW77}. Solid circles represent lower states 
and open squares represent upper states. Solid curves are for $\alpha$=1, 
with best-fit values of the parameter $B$ listed in Table~\ref{tab:fixed1}. 
For $\pm$10\% change of the input single-nucleon amplitudes, the calculated 
fractions change by $\pm$0.4\%. To check sensitivity to fit parameters, 
the long-dashed curves are obtained from equally good fits to kaonic 
atoms data for $\alpha$=2, with $B$ values given by 
\begin{equation} 
B_{\alpha=2}({\rm P})=-0.5+i4.6~{\rm fm}, \,\,\, 
B_{\alpha=2}({\rm KM})=0.3+i3.8~{\rm fm}. 
\label{eq:B} 
\end{equation} 
Estimating the uncertainties arising from the phenomenological part 
$V^{(2)}_{K^-}$ of the optical potential, the calculated fractions 
change by $\pm$4\% upon using extreme values of $\alpha$ between 1 and 2, 
and up to $\pm$0.6\% upon using the results of Table~\ref{tab:fixed1}. 
These estimated uncertainties are comfortably below the experimental 
uncertainty given by the two horizontal dashed lines in Fig.~\ref{fig:PKMfrac}. 

The results for the P and KM sets of amplitudes are indistinguishable on 
the scale of the figure. Indeed, Figs.~\ref{fig:Kpreal} and \ref{fig:Kpimag} 
show that the two sets of amplitudes are very similar to each other over most 
of the energy range between threshold and 30-40~MeV below. This is the 
subthreshold energy range reached by applying the density-to-energy 
transformation Eq.~(\ref{eq:SCmodifMS}) for densities as high as 50\% of 
$\rho_0$. The contribution to the level widths from higher density regions 
is suppressed owing to the poor overlap of the atomic wave functions with 
the nucleus at these densities. 

\begin{figure}[!ht]
\begin{center}
\includegraphics[height=90mm,width=0.80\textwidth]{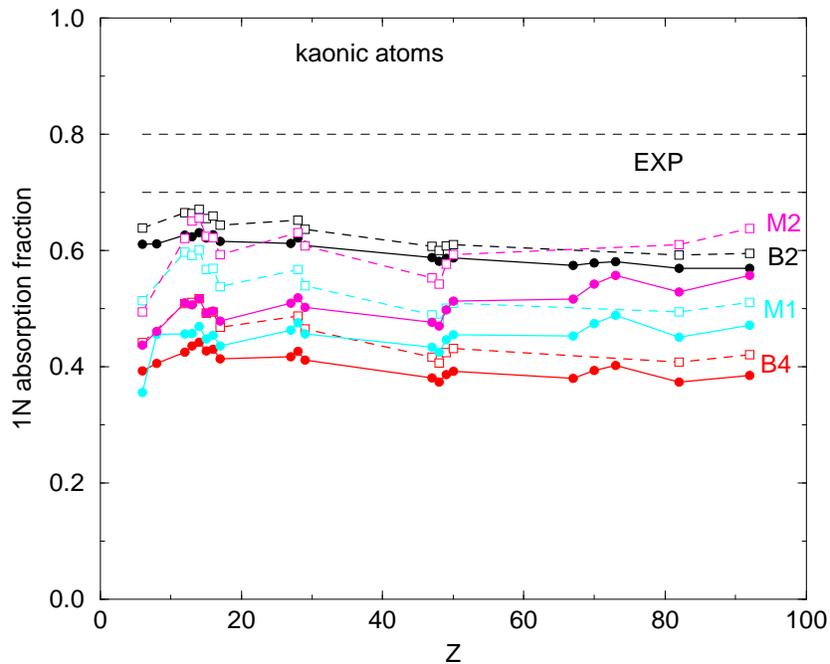}
\caption{Color online. Fractions of single-nucleon absorption for amplitudes 
B2, B4, M1 and M2, using $\alpha$ and $B$ values from Table \ref{tab:fixed1}. 
Solid circles and solid curves are for lower states, open squares and 
long-dashed curves are for upper states. Horizontal dashed lines mark 
the range of experimental values. See text for discussion of uncertainties.} 
\label{fig:BMfrac}
\end{center}
\end{figure} 

In comparing between calculation and experiment we note that for almost all 
species of kaonic atoms in the data base, the absorptions from the upper state 
are of the order of 10-15\% of all absorptions, as deduced from the measured 
upper level to lower level radiation yields. Carbon is an exception with, 
by far, the lowest {\it radiation} yield (of 0.07$\pm$0.013). Atomic cascade 
calculations show that for C about 75\% of the absorptions take place from 
the upper state. Therefore the calculated single-nucleon absorption fractions 
should be close to the upper level points for C and very close to the lower 
level points for all the other species. We conclude that the agreement of 
calculations based on the P and KM amplitudes with the estimated experimental 
range for the single-nucleon absorption fractions is very good.

Figure \ref{fig:BMfrac} shows comparisons between calculations and the 
experimental range for the other four potentials, using $\alpha$ and $B$ 
values listed in Table~\ref{tab:fixed1}. Again, solid circles are for 
lower states and open squares are for upper states. The computed values 
of single-nucleon absorption fractions for these potentials, as shown in 
the figure, fall short significantly of the range of values deduced from 
bubble-chamber experiments. Thus, it is evident that none of these potential 
models is in agreement with experiment.

\section{Discussion}
\label{sec:discuss}

With the construction of in-medium $K^-N$ amplitudes as detailed above, it is 
desirable to re-examine several issues discussed in earlier studies of purely 
phenomenological $K^-$ nuclear optical potentials. In these studies, the 
simplest possible phenomenological `$t\rho $' optical potential of the form 
\begin{equation} 
2 \mu_{K^-} V_{K^-}^{t\rho} = - 4\pi b\,\rho 
\label{eq:trho} 
\end{equation} 
was adopted, with a density independent isoscalar complex amplitude $b$, 
leading to reasonably good fits to kaonic atom data \cite{BFG97,FGa07}. 
However, the fitted constant amplitude $b$ did not reflect any known property 
of the $K^-$ nuclear interaction, and the real and imaginary parts of 
$V_{K^-}^{t\rho}$ in the nuclear interior were merely extrapolations from 
the surface region. Imposing the low-density limit by requiring a slightly 
repulsive threshold amplitude $b^{({\rm th})}$ and adding a phenomenological 
isoscalar density-dependent term, 
\begin{equation} 
b \to b^{({\rm th})} + B (\frac{\rho}{\rho_0})^\alpha, \;\;\;\;\; \alpha >0, 
\label{eq:DD} 
\end{equation} 
improved fits were obtained, associated with compression of the real part of 
the potential \cite{FGB93}. However, the dependence on density, in particular 
the density independence of the single-nucleon term $b^{({\rm th})}$, did 
not result from any microscopic model. Thus, extrapolations to the nuclear 
interior were biased by the simple ansatz (\ref{eq:DD}). In contrast, 
the present formulation of a single-nucleon part $V^{(1)}_{K^-}$ of the 
$K^-$-nucleus optical potential is based on a {\it microscopic} model approach 
to the free-space energy dependent $K^-N$ amplitude and to the way it evolves 
into a single-nucleon in-medium amplitude. Phenomenology enters in our 
construction only by adding a multinucleon part $V^{(2)}_{K^-}$. This approach 
was developed in a series of recent works \cite{CFG11a,CFG11b,FGa12,FGa13}. 

Having chosen within the present approach a preferred 
set of $K^- N$ amplitudes (sets P and KM are essentially equivalent in this 
respect), it is instructive to examine in detail 
the KM {\it in-medium} single-nucleon and 
full amplitudes. This is done in the next two subsections, while in the 
third subsection we remark on the role of `subthreshold kinematics' in 
absorption-at-rest calculations. 

\begin{figure}[!ht] 
\begin{center} 
\includegraphics[height=60mm,width=0.70\textwidth]{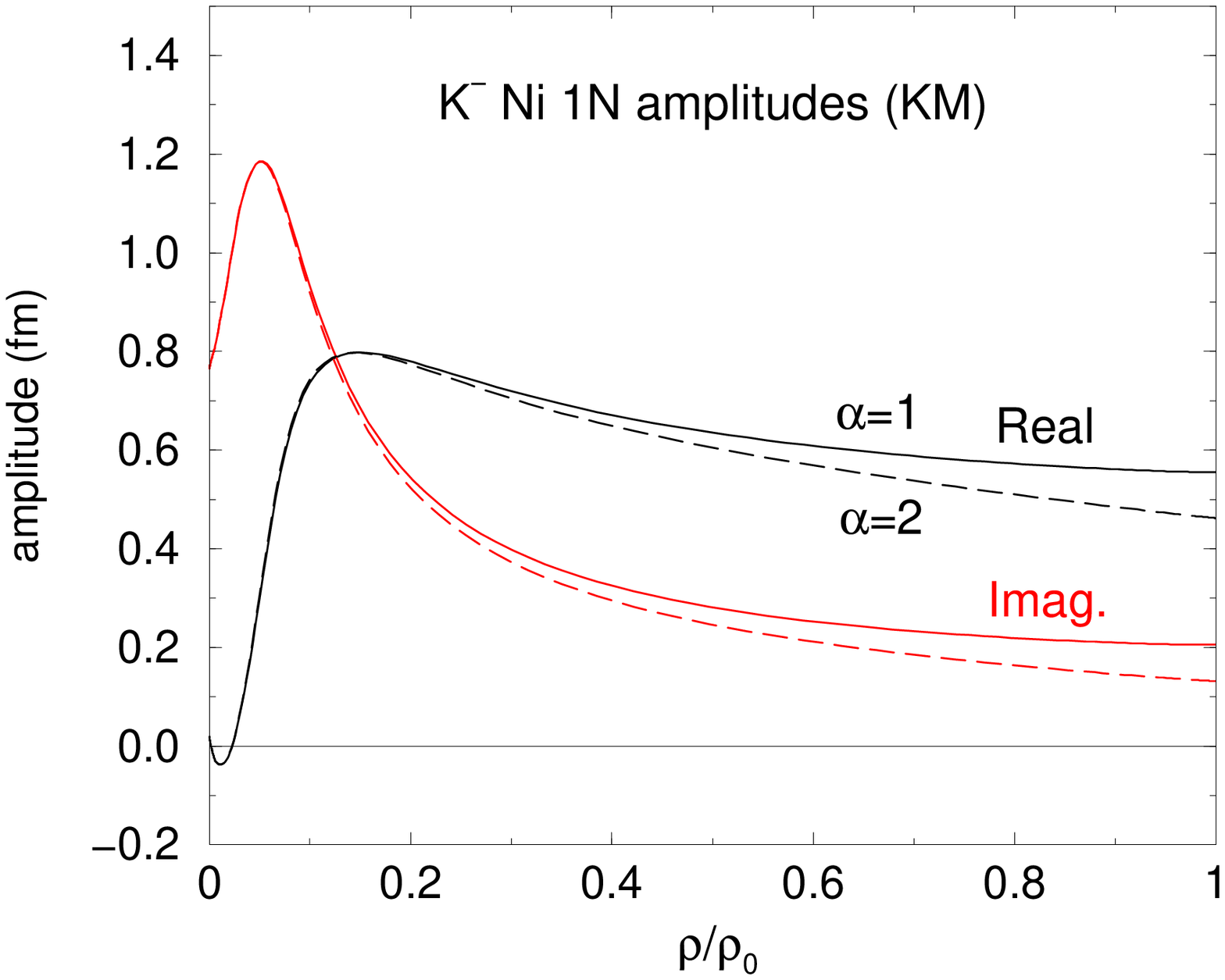} 
\includegraphics[height=60mm,width=0.70\textwidth]{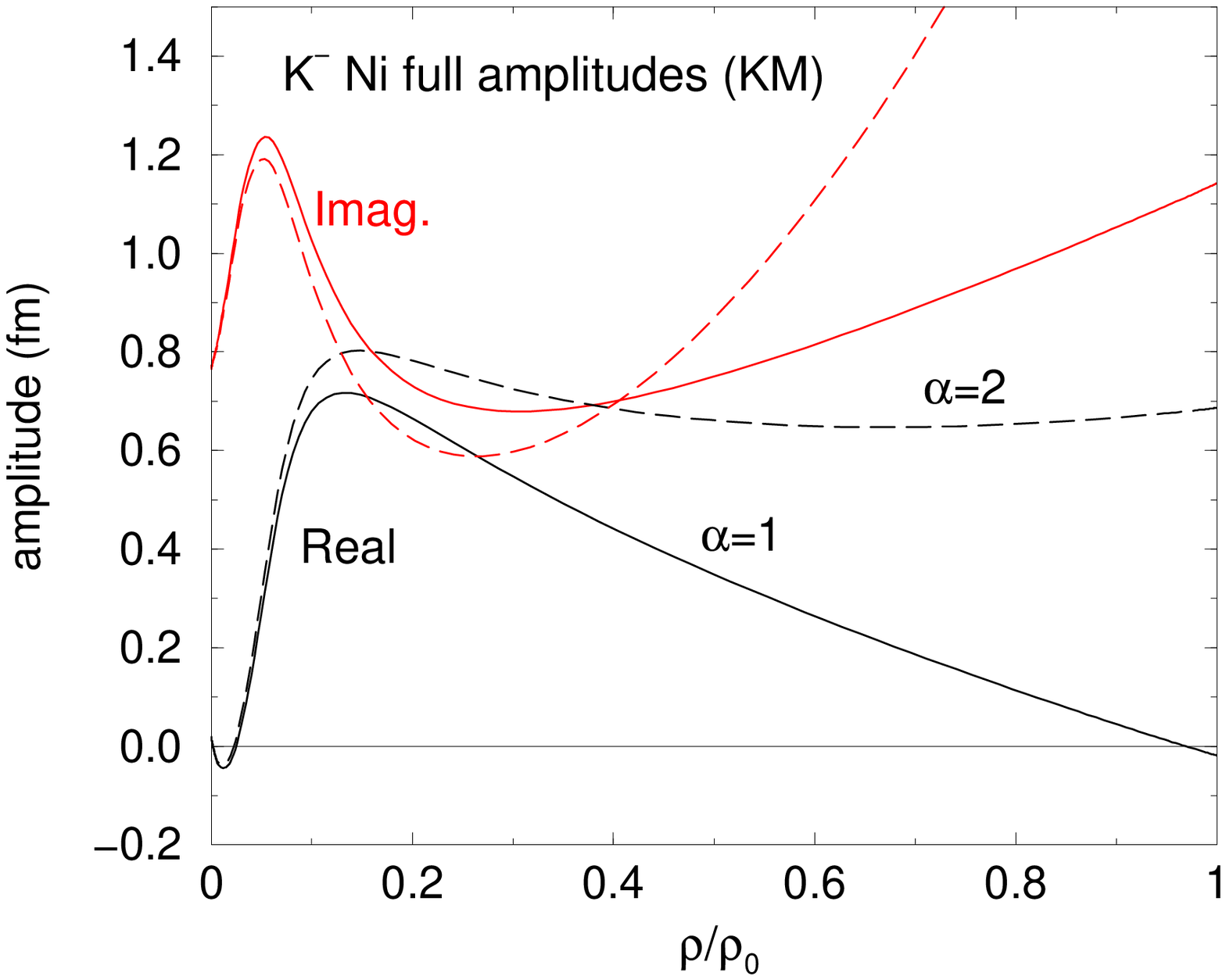} 
\caption{Color online. Upper panel: in-medium single-nucleon $K^- N$ 
amplitudes in Ni from best fits to 65 kaonic-atoms data points, using the 
free-space KM model amplitudes in $V^{(1)}_{K^-}$, Eq.~(\ref{eq:wrw14}), 
plus a phenomenological multinucleon amplitude $B(\rho /\rho_0)^\alpha$ 
in $V^{(2)}_{K^-}$, Eq.~(\ref{eq:V2}), with $\alpha $=1 (solid curves) 
and $\alpha $=2 (dashed curves). Eq.~(\ref{eq:SCmodifMS}) with the full 
$V_{K^-}$ was solved self consistently to obtain the density dependence 
of $V^{(1)}_{K^-}$. Lower panel: the {\it full} in-medium $K^-$ amplitudes 
corresponding to $V_{K^-}=V^{(1)}_{K^-}+V^{(2)}_{K^-}$. See text for
definitions.} 
\label{fig:interplay} 
\end{center} 
\end{figure} 

\subsection{Interplay between theory and phenomenology} 
\label{subsec:interplay} 

Here we discuss the geometry of best-fit $K^-$ nuclear amplitudes and optical 
potentials. Recalling Eq.~(\ref{eq:wrw14}) for $V^{(1)}_{K^-}$, we define the 
expression within the square brackets on the r.h.s. as the in-medium $K^- N$ 
amplitude times $\rho$. Likewise for $V_{K^-}=V^{(1)}_{K^-}+V^{(2)}_{K^-}$, 
the expression $-(2\mu_{K^-}/4 \pi)V_{K^-}(\rho)$ may be defined as the 
full in-medium amplitude times $\rho$. Figure \ref{fig:interplay} shows, 
as an example, in-medium scattering amplitudes as function of the relative 
density in the Ni nucleus. Plotted in the upper panel are in-medium 
single-nucleon $K^- N$ amplitudes, averaged over protons and neutrons, from 
the KM chiral model, for two choices of the multinucleon phenomenological 
exponent $\alpha$. The slight dependence on $\alpha$ observed in this panel 
is due to the dependence of $\delta\sqrt s$ on the full optical potential 
in Eq.~(\ref{eq:SCmodifMS}). 
Plotted in the lower panel are full in-medium amplitudes, including the 
added phenomenological term $B(\rho /\rho_0)^\alpha$. It is seen that up to 
a density of 40--50\% of central density $\rho_0$ the two values of $\alpha$ 
lead to about the same imaginary parts of the amplitudes (and potentials). 
Recall that both values of $\alpha$, and the associated complex values of 
the parameter $B$, were obtained from best fits to the kaonic atoms data. The 
model dependence of the real part of the full in-medium amplitude is stronger, 
reflecting the fact that for kaonic atoms the imaginary part of the potential 
is dominant. In fact, strong interaction level widths are significantly 
larger than the corresponding level shifts in kaonic atoms. Clearly, 
the full in-medium amplitudes (and potentials) for densities greater 
than 40--50\% of central density are just analytical continuation of 
whatever parameterization was employed for the phenomenological terms. 

Systematics of purely phenomenological $K^- $-nucleus optical potentials 
showed \cite{FGB93} that the r.m.s. radius of the real part was usually 
smaller than the corresponding radius of the density distribution of the 
nucleus. This can be attributed to the sharp rise of the real part of the 
full in-medium amplitude with increased density near the nuclear surface. 
Figure~\ref{fig:interplay} shows that a similar rise of the imaginary part 
of the full in-medium amplitude is followed, still in the surface region, 
by a decrease as the density increases, which could cause the reverse trend 
in the r.m.s. radius. Indeed, phenomenological potentials show that the r.m.s. 
radius of the imaginary part can be either somewhat larger or smaller than 
that of the nucleus.

\subsection{Radial sensitivity of kaonic atoms}
\label{subsec:radial}

\begin{figure}[!ht]
\begin{center}
\includegraphics[height=65mm,width=0.75\textwidth]{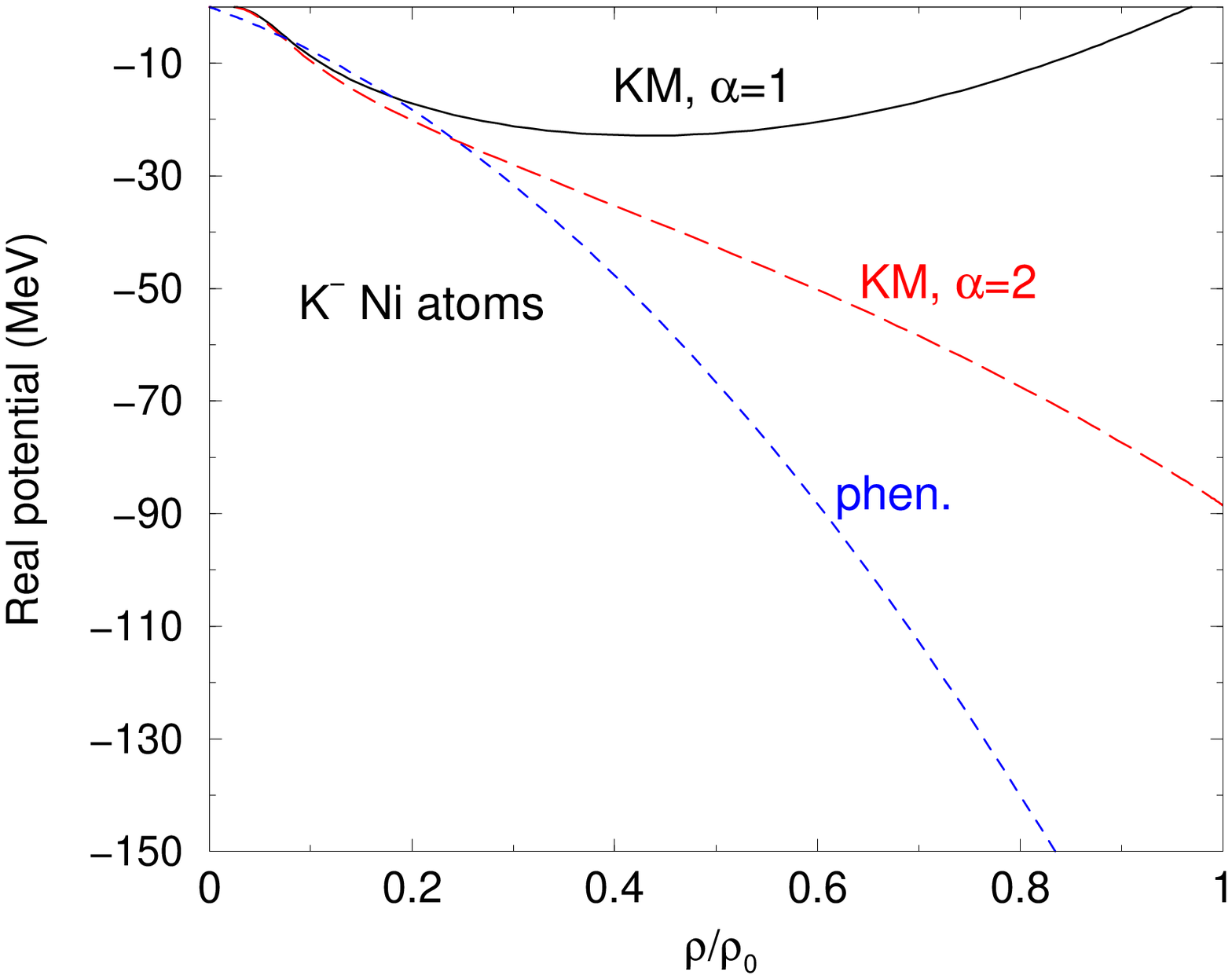}
\includegraphics[height=65mm,width=0.75\textwidth]{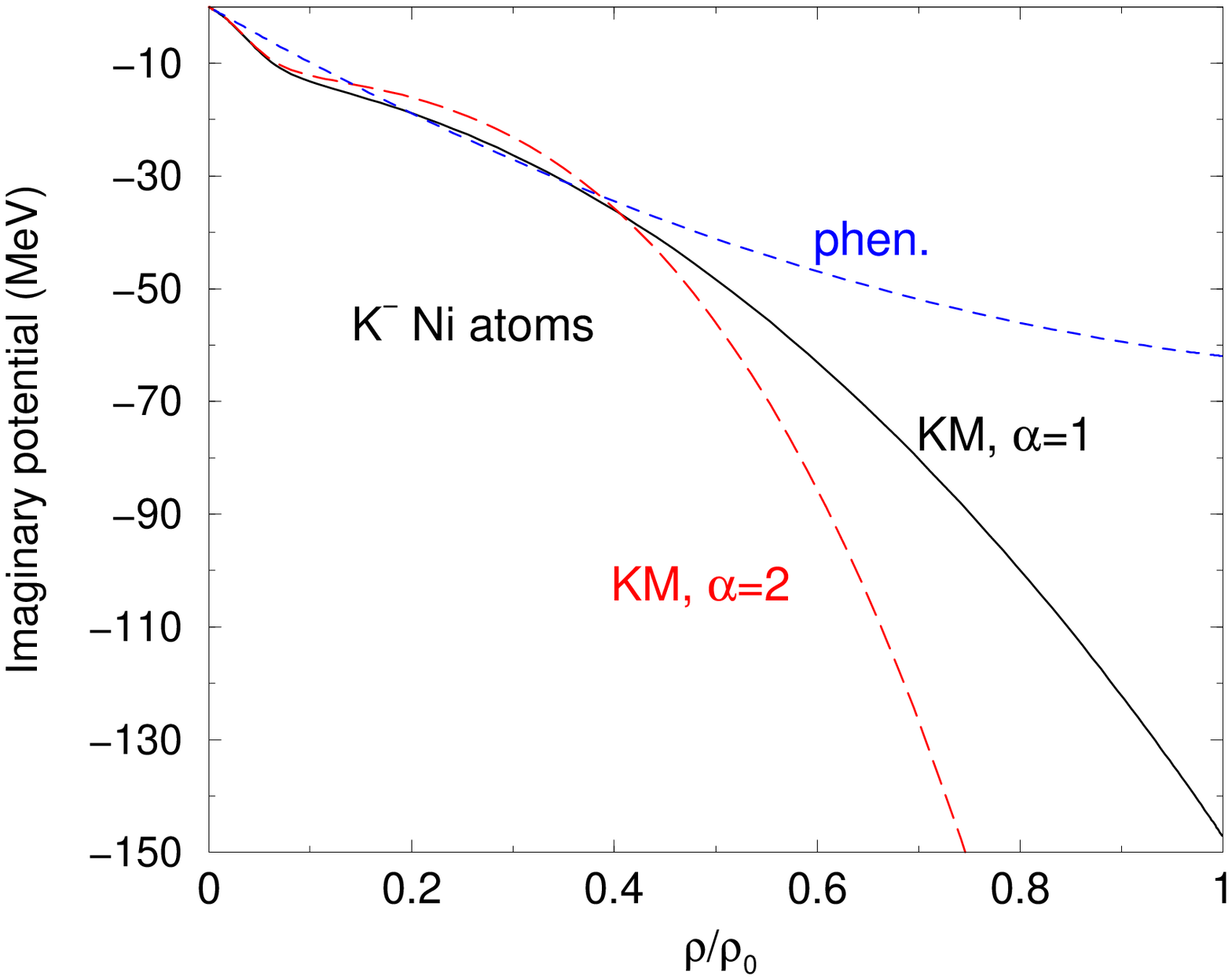}
\caption{Color online. Real part (upper panel) and imaginary part
(lower pannel) of best-fit $K^-$ optical potentials for kaonic atoms of Ni.
Solid curves and long-dashed curves are based on the KM single-nucleon
amplitudes plus a phenomenological term $B(\rho /\rho_0)^\alpha$, dashed
curves are for a purely phenomenological density-dependent best-fit
potential \cite{FGa07}. All three potentials lead to equally good fits
to the 65 data points in kaonic atoms.}
\label{fig:radial}
\end{center}
\end{figure}

The issue of radial sensitivity of kaonic atom experiments is examined next. 
Figure \ref{fig:radial} shows, as an example, three versions of optical 
potentials for Ni that fit equally well the whole set of kaonic atoms data 
across the periodic table. The KM potentials for $\alpha $=1 and $\alpha$=2 
consist of a single-nucleon potential $V^{(1)}_{K^-}$ (\ref{eq:wrw14}) 
based on the KM $K^-N$ amplitudes, augmented by a phenomenological potential 
$V^{(2)}_{K^-}$ (\ref{eq:V2}) with a complex parameter $B$ listed in 
Table~\ref{tab:fixed1} for $\alpha$=1 and in Eq.~(\ref{eq:B}) for $\alpha$=2. 
Both of these KM versions of $V_{K^-}$ satisfy the low-density limit and 
reproduce well the experimentally based single-nucleon absorption-at-rest 
fractions considered in the present work. In contrast, the third potential, 
a purely phenomenological density-dependent potential of the form 
(\ref{eq:DD}) with $\alpha$=1, does not necessarily respect any of these 
constraints. The potentials are plotted as function of the density, relative 
to nuclear matter density $\rho_0$. It is seen that for the imaginary part 
of the potential all three curves almost coincide up to 40\% of $\rho_0$, 
and only above 50\% of $\rho_0$ they begin to depart from each other. The 
real parts differ from each other already above 30\% of $\rho_0$. The wild 
disagreement between these models for densities larger than 60\% of $\rho_0$, 
in spite of all three potentials producing equally good fits to the data, 
indicates that kaonic atoms are totally insensitive to these density regions 
in nuclei. This is in line with the overlaps presented in 
Fig.~\ref{fig:overlaps} for the atomic wave functions and the imaginary part 
of the potential. The strength of the well-defined imaginary part of the 
potential in the nuclear surface region ensures damping of the atomic wave 
function when the density exceeds about 50\% of $\rho_0$. The potential 
in the surface region is dominated by the single-nucleon amplitudes that 
are firmly rooted in theory and are validated here, upon introducing 
a best-fit phenomenological potential $V^{(2)}_{K^-}$, by the single-nucleon 
absorption fraction constraint.

As noted above, the behavior of Re~$V_{K^-}(\rho)$ at densities beyond 
$\approx 0.2\,\rho_0$, as shown in the upper panel of Fig.~\ref{fig:radial}, 
represents merely extrapolation to a density regime which is uncontrolled by 
fitting to kaonic atoms data. Yet, one cannot avoid the question of why the 
resulting depth of Re~$V_{K^-}(\rho)$ at nuclear matter density $\rho_0$ 
in the two KM versions shown is so much smaller than that produced in the 
purely phenomenological model \cite{FGa07}, also shown in the figure, 
and in more recent analyses based on single-nucleon $\bar K N$ chiral 
approaches \cite{FGa12,FGa13}. To answer this query we note that the 
added $V^{(2)}_{K^-}$ in the latter works consisted of two terms, one linear 
and the other approximately quadratic in the nuclear density. The linear term 
was interpreted as possibly correcting the the single-nucleon input of 
$V^{(1)}_{K^-}$, and it came out {\it repulsive} by fitting to the data 
whereas the quadratic term came out {\it attractive}. At $\rho\lesssim 
0.2\rho_0$, where the fit is still meaningful, the overall sign of 
$V^{(2)}_{K^-}$ was determined by the repulsive linear term, but at larger 
densities the quadratic term took over and made $V^{(2)}_{K^-}$ attractive. 
In the present work we insist on respecting the low density limit which 
rules out including a linear (in the density) term beyond $V^{(1)}_{K^-}$. 
The role of this linear term in the previous analyses \cite{FGa12,FGa13} 
is now taken over by the only multinucleon term represented by the complex 
$B$ parameter, so it now comes {\it repulsive} in most fits. This example 
demonstrates explicitly how the high density behavior of Re~$V_{K^-}(\rho)$ 
is no more than just extrapolation based on whatever parametrization, 
valid at the low-density regime, one uses in the analysis.

\subsection{Role of `subthreshold kinematics' in absorption-at-rest 
calculations}
\label{subsec:jido} 

The present calculations of single-nucleon absorption fractions make use of 
the `subthreshold kinematics' formulation to evaluate self consistently the 
in-medium single-nucleon part $V^{(1)}_{K^-}(\rho)$ of the $K^-$ nuclear 
optical potential $V_{K^-}$ applicable in kaonic atoms. In this formulation, 
the downward energy shift $\delta\sqrt{s}$ at which the free-space $K^- N$ 
scattering amplitudes need to be evaluated in the nuclear medium, translates 
into a density dependence of the in-medium scattering amplitudes. A major 
contribution to this $\delta\sqrt{s}$, Eq.~(\ref{eq:SCmodifMS}), arises 
from the nonvanishing in-medium kaon momentum $p_{K^-}$, related in the 
local density approximation to $V_{K^-}$ as given by Eq.~(\ref{eq:locpi}). 
We recall that a $K^-$ meson orbiting in kaonic atoms is not at rest when 
exposed to the strong-interaction nuclear force. Here we wish to demonstrate 
the significant effect of this kaon-momentum contribution on the $K^-$ 
absorption fractions calculated in Sect.~\ref{subsec:1Nfrac} and compare with 
other calculations that disregard it. 

\begin{table}[htb]
\caption{Best-fit values of Re$\,B$ and Im$\,B$ for $\alpha$=1, 
and single-nucleon fractions of absorption from the lower and upper 
states in the kaonic Ni atom, for three model amplitudes used  
 to evaluate $V_{K^-}(\rho)$.} 
\label{tab:1Na} 
\begin{center} 
\begin{tabular}{cccccc} 
\hline 
amplitude & $\chi ^2$(65) & Re$B$ (fm) & Im$B$ (fm) & lower & upper \\  \hline 
KM & 122 & $-$0.9$\pm$0.2 & 1.4$\pm$0.2 & 0.74 & 0.81 \\ 
P& 125 & $-$1.3$\pm$0.2 & 1.5$\pm$0.2 & 0.73 & 0.80 \\ 
M2 & 109 & 2.1$\pm$0.2 & 1.2$\pm$0.2 & 0.52 & 0.63 \\  \hline 
\end{tabular} 
\end{center} 
\end{table} 

Table \ref{tab:1Na} lists single-nucleon absorption fractions for the 
lower and upper states in kaonic Ni atoms, as calculated in the present 
work and shown in Figs.~\ref{fig:PKMfrac} and \ref{fig:BMfrac}, using the 
three model amplitudes that allow in their best fits a multinucleon term 
$V^{(2)}_{K^-}(\rho)$ with $\alpha$=1. Since most of the absorption in kaonic 
Ni occurs from the lower state, the table demonstrates that models KM and P 
reproduce the experimentally deduced single-nucleon absorption fraction of 
0.75$\pm$0.05, whereas model M2 fails to do so by a wide margin. 

\begin{table}[htb]
\caption{Same as in Table \ref{tab:1Na}, except that $\delta\sqrt{s}$ of 
Eq.~(\ref{eq:SCmodifMS}) is modified by assuming that the in-medium $K^-$ 
momentum $p_{K^-}$ is zero, see text.}
\label{tab:1Nb}
\begin{center}
\begin{tabular}{cccccc}
\hline
amplitude & $\chi ^2$(65) & Re$B$ (fm) & Im$B$ (fm) & lower & upper \\  \hline
KM & 140 & $-$1.3$\pm$0.2 & 0.8$\pm$0.2 & 0.85 & 0.89 \\ 
P & 141 & $-$1.6$\pm$0.2 & 0.9$\pm$0.2 & 0.84 & 0.89 \\ 
M2 & 159 & 0.0$\pm$0.1 & 1.3$\pm$0.2 & 0.67 & 0.76 \\  \hline 
\end{tabular} 
\end{center} 
\end{table} 

Table \ref{tab:1Nb} is constructed similarly to Table \ref{tab:1Na}, 
but the  contribution of the kaon momentum $p_{K^-}$ to the r.h.s. of 
Eq.~(\ref{eq:SCmodifMS}) is switched off. The $\chi ^2$ values of these 
best-fits are not as good as those listed in Table \ref{tab:1Na} and 
the single-nucleon fractions calculated in models KM and P no longer 
reproduce the experimental ones. The general trend of the calculated 
single-nucleon absorption fractions is to increase with this modification 
of $\delta\sqrt{s}$ in Eq.~(\ref{eq:SCmodifMS}). This still leaves model M2 
unacceptable as far as absorption from the dominant lower state  
is concerned, although it brings the fraction calculated for absorption 
from the upper state into accordance with experiment. 
Note that the values of $\delta\sqrt{s}$ for $p_{K^-}=0$ are smaller by 
about 10 MeV with respect to those obtained with full account of $p_{K^-}$ 
according to Eq.~(\ref{eq:locpi}). The variation of Im$\,V_{K^-}$ over 
this energy interval is sufficiently strong  to cause this increase 
in the calculated single-nucleon absorption fractions. 

Finally, we comment on the nuclear-matter calculation of $K^-$ absorption 
fractions in Ref.~\cite{SYJ12}, in particular those 
shown in Fig. 17. In this work a microscopic model, 
using meson-exchange contributions dominated by the subthreshold 
$\Lambda ^{\ast}$(1405) resonance, was formulated for the absorptive 
parts of {\it both} $V^{(1)}_{K^-}(\rho)$ and $V^{(2)}_{K^-}(\rho)$. 
The single-nucleon absorption fraction decreases in this calculation from 
the value 1 for $\rho\to 0$, reaching the experimentally deduced value 
of 0.75$\pm$0.05 at a density about 70--90\% of $\rho_0$, which is too 
high value compared to where absorption takes place in kaonic atoms. 
At typical kaonic atoms absorption densities of 30--40\% of $\rho_0$, 
the single-nucleon absorption fraction calculated in this model is about 
0.85 to 0.90. This should decrease, according to the insight gained by 
comparing the absorption fractions listed in Table \ref{tab:1Nb} 
with those in 
Table \ref{tab:1Na}, once the assumption of $K^-$ at rest is relaxed in 
the calculations of Ref.~\cite{SYJ12}, perhaps offering a better agreement 
with experiment.

\section{Summary}
\label{sec:summ}

Six recent sets of $K^-N$ amplitudes near threshold, based on 
SU(3) chiral-model EFT approaches to the $\bar K$-nucleon interaction, 
have been tested as basic components in the construction of $K^-$-nucleus 
optical potentials that fit kaonic atom strong-interaction data across 
the periodic table. All sets need to be supplemented by a phenomenological 
term in order to fit the data. However, imposing as additional constraint 
the single-nucleon fraction of absorption at rest, only two of the six 
$K^-N$ model amplitudes are found capable of reproducing the absorption 
fractions obtained from bubble-chamber experiments. The two acceptable 
models, marked as KM for Kyoto-Munich and P for Prague, produce very 
similar in-medium $\bar KN$ amplitudes down to about 40 MeV below threshold, 
which is the region of energies relevant for kaonic atoms.

\section*{Acknowledgements}

We thank Ale\v{s} Ciepl\'{y} for providing us with $K^- N$ amplitudes from 
Ref.~\cite{CMM16} in tabulated form. Useful remarks by him and by Ji\v{r}\'{i} 
Mare\v{s} on the penultimate version of this work are gratefully acknowledged.

\end{document}